\documentclass[fleqn,twoside]{article}

\usepackage{amsmath}
\usepackage{amsfonts}
\usepackage{amssymb}
\usepackage{graphicx}%
\def\be{\begin{equation}}
\def\ee{\end{equation}}
\def\bi{\bibitem}

\begin{document}
\title{Cosmological Lorentzian Wormholes via Noether symmetry approach}
\author{Abhik Kumar Sanyal$^\dag$ and Ranajit Mandal$^\ddag$, }
\maketitle \noindent
\begin{center}
\noindent
$^{\dag}$ Dept. of Physics, Jangipur College, Murshidabad,
\noindent
West Bengal, India - 742213. \\
$^{\ddag}$ Dept. of Physics, University of Kalyani, West Bengal, India - 741235.\\
\noindent

\end{center}
\footnotetext[1] {\noindent
Electronic address:\\
\noindent
$^{\ddag}$sanyal\_ ak@yahoo.com;
$^{\ddag}$ranajitmandalphys@gmail.com\\}

\begin{abstract}

Noether symmetry has been invoked to explore the forms of a couple of coupling parameters and the potential appearing in a general scalar-tensor theory of gravity in the background of Robertson-Walker space-time. Exact solutions of Einstein's field equations in the familiar Brans-Dicke, Induced gravity and a General non-minimally coupled scalar-tensor theories of gravity have been found using the conserved current and the energy equation, after being expressed in a set of new variables. Noticeably, the form of the scale factors remains unaltered in all the three cases and represents cosmological Lorentzian wormholes, analogous to the Euclidean ones. While classical Euclidean wormholes requires an imaginary scalar field, the Lorentzian wormhole do not, and the solutions satisfy the weak energy condition.
\end{abstract}
\noindent
\section{\bf{Introduction}}

Apart from Black holes, Wormholes are yet another extraordinary, exciting and intriguing consequence of Einstein's General Theory of Relativity (GTR) $G_{\mu\nu} = R_{\mu\nu} - {1\over 2}g_{\mu\nu} R = \kappa T_{\mu\nu}$. Since the pioneering works of Lavrelashvili, Rubakov and Tinyakov \cite{1, 2, 3} followed by Giddings and Strominger \cite{4} and thereafter by Morris and Thorne \cite{5,6}, wormholes turn out to be one of the most popular and intensively studied topics in Astronomy. Wormholes are essentially astrophysical objects which connect two asymptotically flat or de-Sitter/ anti-de-Sitter regions by a throat of finite radius. While, microscopic wormholes might provide us with the mechanism that might be able to solve the cosmological constant problem, macroscopic wormholes might be responsible for the final stage of evaporation and complete disappearance of black holes. Since the Wheeler-DeWitt equation is independent of the lapse function and as such holds for both the Euclidean and Lorentzian geometry, so following Hawking and Page formulation \cite{7} Euclidean wormholes may be realized in the early universe, but only for some specified forms of the scalar potentials \cite{8}. That is why, most of the efforts have been directed to the study of Lorentzian wormholes in the framework of classical GTR. The striking feature of wormholes is the requirement of the violation of energy conditions. Therefore, realization of Lorentzian wormhole solutions with standard barotropic fluid is not acceptable physically. This implies that the matter supporting the traversable wormholes (wormholes without a horizon) should be exotic \cite{5,6,9,10,11}. and therefore it should have very strong negative pressure, or even that the energy density may be negative. Therefore, a lot of efforts have been directed to the study Lorentzian wormholes, in the framework of classical general relativity, sustained by an exotic matter with negative energy density. In general, these models include both static \cite{12,13,14,15,16,17,18,19,20,21,22} and evolving relativistic versions \cite{23,24,25,26,27,28,29,30,31,32,33}, sustained by a single fluid component. The interest has been mainly devoted to the study of traversable wormholes, without any horizon, allowing two-way passage through them \cite{34}. For static wormholes the fluid requires the violation of the null energy condition (NEC), while in Einstein gravity there exists non-static Lorentzian wormholes which do not require weak energy condition (WEC) violating matter to sustain them. Such wormholes may exist for arbitrarily small or large intervals of time \cite{23,24}, or even satisfy the dominant energy condition (DEC) in the whole spacetime \cite{35,36}.\\

Scalar fields may be treated as the most common candidates for wormholes, with such exotic behaviour. At this point it is important to understand that GTR can accommodate all sorts of matter fields through the energy-momentum tensor $T_{\mu\nu}$. In this sense, scalar-tensor theory of gravity in principle should not be treated as a modification of GTR. Nevertheless, in view of the action principle, non-minimally coupled scalar-tensor theory of gravity may be looked upon as a modification of GTR, since it requires coupling between the Ricci scalar $R$ and some arbitrary function $f(\phi)$ of the scalar field $\phi$ in the form $f(\phi) R$, in the action. But in view of the field equations, it might just again be treated as incorporating a typically different energy-momentum tensor altogether. Only by modifying Einstein-Hilbert action by introducing different higher-order curvature invariant terms, left hand side of the Einstein's equation and hence GTR is truly modified. In this sense wormhole solutions for scalar-tensor theory of gravity may also be treated as a consequence of GTR. In the context of cosmology, the violation of energy condition for such matter fields in the early universe, does not in any way affect the late stage of cosmic acceleration. \\

Evolving Lorentzian wormholes in the background of Robertson-Walker metric has already been studied by some authors \cite{36, 37, 38, 39}. In general, while constructing wormhole geometries, first the form of the redshift function $\Phi(r)$ and the shape function $b(r)$, satisfying some general constraints \cite{5,6} are fixed. This fixes the metric is as well. Thereafter in view of the field equations, components of the energy-momentum tensor required to support the spacetime geometry, are explored. For evolving wormholes, one usually generalizes the ansatz for static Lorentzian wormhole given by Morris and Thorne \cite{5,6} in the form,

\be ds^2 = -e^{2 \Phi(r,t)}dt^2+a(t)^2\left[\frac{dr^2}{1-{b(r,t)\over r}}+r^2(d\theta^2+sin^2\theta\; d\phi^2)\right].\ee
However, the evolving Lorentzian wormhole, we are going to study in the present manuscript is different altogether. We follow the same standard method of solving Einstein's equations, as usually employed in the case of Euclidean wormholes. That is, we do not fix the redshift function $\Phi(r,t)$ or the shape function $b(r,t)$, and the reason is, shape of a cosmological wormhole is fixed by the scale factor itself \cite{8, 40, 41, 42, 43}. Practically, our initial aim is just to study the evolution of the early universe in view of the non-minimally coupled scalar-tensor theory of gravity, corresponding to the action being typically expressed in the following form,

\be\label{Action}
A = \int d^{4}x \sqrt{-g} \bigg{[}\frac{f(\phi)R-2\Lambda}{16\pi G}-\omega(\phi)\frac{1}{2}\phi_{,\mu}\phi^{,\mu}-V(\phi)\bigg{]}, \ee
in the background of isotropic and homogeneous Robertson-Walker metric,

\be\label{RW} ds^2 = -dt^2+a(t)^2\left[\frac{dr^2}{1-kr^2}+r^2(d\theta^2+sin^2\theta\; d\phi^2)\right].\ee
Note that since we are interested in the evolution of early universe, so we ignore any form of baryonic matter what-so-ever. The action \eqref{Action} involves the cosmological constant $\Lambda$, and so is a generalization of our earlier work \cite{44}. It also involves two coupling parameters $f(\phi)$ and $\omega(\phi)$ apart from the potential $V(\phi)$, whose forms are required, to explore the evolution. Instead of choosing the forms of these parameters by hand, we apply Noeher symmetry for the purpose. In view of the forms of the parameters, the solutions so obtained have been found to represent wormholes in the sense that asymptotically ($t\rightarrow \pm \infty$), the scale factor ($a(t)$) tends to de-Sitter universe, while for $t \rightarrow 0$, it is finite, which is the throat of the wormhole. We call it cosmological wormhole, since the universe evolves like a wormhole and it initiates inflation.\\

In general, Noether symmetry plays an important role in physics because it can be used to simplify a given system of differential equations as well as to determine the integrability of the system. Due to the Noether theorem, symmetries are always related to conserved quantities which, in any case, can be considered as conserved ‘charges’. Specifically, the form of the self interacting scalar-field potential is ‘selected’ by the existence of a symmetry and then the dynamics can be controlled. In Search of Noether symmetries in the Robertson-Walker space-time background, first it is required to construct the point Lagrangian from the general non-minimally coupled gravitational action \eqref{Action}. Numerous works regarding the study of Noether symmetry for scalar-tensor theory of gravity is available in the literature since its inception \cite{45, 46}. However, there is a typical problem associated with it, which has often been overlooked. In the following section we describe the basic ingredient of Noether symmetry and the associated problem. In section 4, we shall explore the symmetries in three different situations for different forms of $f(\phi)$, viz. the `Brans-Dicke theory', the `Induced theory of gravity' and the `General non-minimally coupled scalar-tensor theory of gravity', associated with different forms of the coupling parameter $\omega(\phi)$ and the potential $V(\phi)$. In all the cases the scale factor retains the same form whose nature admits Lorentzian wormholes, which don't violate WEC. It thus appears that Lorentzian wormholes are almost a generic feature of the system \eqref{Action} under consideration. We conclude in section 5.

\section{\bf{ Basic ingredients of Noether symmetry}}

Noether symmetry approach is a powerful tool in finding unknown parameters e.g. the potential and the coupling parameters appearing in the Lagrangian. Using this method it is possible to obtain a reduction of the field equations and sometimes to obtain a full integration of the system, once the cyclic variable of the system is found. The key point related to the Noether symmetry is a Lie algebra presented in the tangent space. Noether theorem states that, if there exists a vector field ${X}$, for which the Lie derivative of given Lagrangian $L$ vanishes, i.e. $\pounds_X L=0$, the Lagrangian admits a Noether symmetry and thus yields a conserved current. If the Lagrangian under consideration is spanned by the configuration space $M = (a,\phi)$, then the corresponding tangent space is TM = ($a,\phi,\dot a,\dot \phi$). Hence the generic infinitesimal generator of the Noether symmetry is

\be\begin{split}\label{X} X & = \alpha(a,\phi)\frac{\partial}{\partial a}+\beta(a,\phi)\frac{\partial}{\partial \phi}+\dot \alpha(a,\phi)\frac{\partial}{\partial\dot a}+\dot\beta(a,\phi)\frac{\partial}{\partial \dot \phi}\\&
\mathrm{where,}~~\dot\alpha\equiv\frac{\partial\alpha}{\partial a}\dot a+\frac{\partial\alpha}{\partial \phi}\dot\phi;\; \dot\beta\equiv\frac{\partial\beta}{\partial a}\dot a+\frac{\partial\beta}{\partial \phi}\dot \phi.\end{split}\ee
In the above, $\alpha,\beta$ are both generic functions of $a$ and $\phi$ and the Lagrangian is invariant under the transformation $X$ i.e.

\be\label{Lx} \pounds_X L = XL=\alpha\frac{\partial L}{\partial a}+\beta\frac{\partial L}{\partial \phi}+\dot \alpha\frac{\partial L}{\partial\dot a}+\dot\beta\frac{\partial L}{\partial \dot \phi} = 0,\ee
where $\pounds_X L$ is the Lie derivative of the point Lagrangian with respect to X. The above equation may be solved to find the unknown parameters of the theory e.g. the potential. However, if there are several unknown parameters, relations connecting those are found, which may be used to select the form of the parameters from physical argument. Now, in view of the Cartan's one form

\be\label{TL} \theta_L=\frac{\partial L}{\partial\dot a}da+\frac{\partial L}{\partial \dot\phi}d\phi,\ee
the constant of motion $Q = i_X\theta_L$, which is essentially the conserved current, is expressed as

\be\label{Q} Q=\alpha(a,\phi)\frac{\partial L}{\partial\dot a}+\beta(a,\phi)\frac{\partial L}{\partial\dot\phi}.\ee
It is well known that the cyclic variable helps a lot in exploring the exact description of the dynamical system. So, once $X$ is found in view of the solutions of the Noether equation \eqref{Lx}, it is possible to change the variables to $u(a,\phi)$ and $v(a,\phi)$, such that $i_X du=1;\; i_Xdv=0$ i.e.

\be\label{cyclic} i_Xdu=\alpha\frac{\partial u}{\partial a}+\beta\frac{\partial u}{\partial\phi}=1;\;\;\; i_Xdv=\alpha\frac{\partial v}{\partial a}+\beta\frac{\partial v}{\partial\phi}=0.\ee
The Lagrangian when expressed in term of the new variables, $u$ becomes the cyclic variable, and the constant of motion $Q$ is its canonically conjugate momentum, i.e. $Q=P_u$. Thus the conserved current assumes a very simple form so that exact integration is often found leading to exact solution of the filed equations under consideration.\\

Nevertheless, as mentioned in the introduction, there is an insidious problem associated with Noether symmetric solutions for gravitational system in particular. Noether symmetry is not on-shell for constrained system like gravity. This means, in general it neither satisfy the field equations nor its solutions by default. Due to diffeomorphic invariance gravity constrains the Hamiltonian (it also constrains the momenta in general, whenever time-space components exists in the metric) of the system to vanish, i.e. the total energy ($E$) of a gravitating system is always zero. This is essentially the ($^0_0$) component of the Einstein's field equations, when expressed in terms of configuration space-variable. Noether equation does not recognize the said constraints (energy and momenta). Therefore, except in some particular case, often it does not satisfy the ($^0_0$) component of Einstein's field equations \cite{47, 48, 49}. Also, sometimes it leads to degeneracy in the Lagrangian \cite{50, 51}. Further, often it fails to explore the known symmetries of the system \cite{52, 53}. Finally, different canonical forms (the point Lagrangian obtained through Lagrange multiplier method, and under reduction to Jordan's and Einstein's frames) of $F(R)$ theory of gravity yield different conserved currents \cite{54}. A possible resolution to the problems \cite{55} is not to fix the gauge, viz. the lapse function $N,$ a-priori, but to keep it arbitrary, so that the ($^0_0$) component of Einstein's equation is recognized by Noether equation. In the process, the symmetry generator $X$ should be modified to

\be\label{gen1} X = \alpha \frac{\partial}{\partial {a}}+\beta\frac{\partial }{\partial \phi} + \gamma \frac{\partial }{\partial {N}} +\dot\alpha \frac{\partial }{\partial\dot {a}}+\dot\beta \frac{\partial }{\partial \dot{\phi}} +\dot\gamma \frac{\partial }{\partial\dot {N}}.\ee
Likewise, when the metric would contain time-space components, one should keep the shift vector ($N_i$) arbitrary. However, with the introduction of lapse function (and shift vector as well) the Hessian determinant vanishes and the point Lagrangian becomes singular \cite{56}. It is therefore required to follow Dirac's constraint analysis, which complicates the problem. Under such circumstances, another proposal has been placed very recently and that is to modify the existence condition for Noether symmetry as \cite{57},

\be \pounds_X L - \eta E - \sum_i \delta_i P_i = 0,\ee
where, $E$ is the total energy of the gravitating system which is constrained to vanish, and essentially is the ($^0_0$) component of Einstein's equations. $P_i$ are the momenta which are also constrained to vanish, and are the $(^0_i)$ components of Einstein's field equations. In the above, $i$ runs from $1$ through to $3$ and $\eta = \eta(a,\phi)$ and $\delta_i = \delta_i(a,\phi)$ are generic functions of $a$ and $\phi$. It has been possible to remove the problems associated with $F(R)$ theory of gravity \cite{57} and Einstein's field equations are automatically satisfied in all circumstances.\\

In the present manuscript however, we take a different route to explore Noether symmetry of the action \eqref{Action}, which has been used earlier \cite{44} and found to be a very powerful technique to find explicit solutions of the field equations. First, we use the standard symmetry generator \eqref{X} and consequently the standard Noether equation \eqref{Lx} to find Noether solutions. Since, there are five unknown parameters of the theory \eqref{Action}, viz. $a, \phi, f(\phi), \omega(\phi), V(\phi)$, consequently there are five unknown parameters $\alpha(a,\phi), \beta(a,\phi), f(\phi), \omega(\phi), V(\phi)$ involved in four Noether equations. So, Noether equations when solved would lead to relations amongst the parameters. To find the forms explicitly, we therefore would require to make yet another choice. We shall therefore assume known standard forms of $f(\phi)$, and solve Noether equations to explore the forms of $\omega(\phi)$ and $V(\phi)$ and consequently $\alpha(a, \phi)$ and $\beta(a, \phi)$. We then express the scale factor $a$, the scalar field $\phi$, the Lagrangian, the conserved current $Q$ and the energy equation, viz. the ($^0_0$) equation of Einstein in terms of the new variables $u$ and $v$. Next, we solve for $u$ and $v$ in view of the last two equations, viz. the conserved current and the energy equations, and transform back to find the explicit solutions for $a(t)$ and $\phi(t)$. Since, ($^0_0$) equation is used for the purpose, so all the Einstein's equations are automatically satisfied. We study three different cases corresponding to three different physical choices of the parameter $f(\phi)$, viz. the `Brans-Dicke form ($f(\phi) = \phi$)', the `Induced gravity theory ($f(\phi) = \epsilon \phi^2$)' and the `General non-minimal coupled theory ($f(\phi) = 1-\epsilon \phi^2$)' and surprisingly observe that all the three cases lead to the same forms of the scale factor ($a(t)$), which are Lorentzian wormhole solutions. It therefore appears that Lorentzian wormhole is a natural outcome of non-minimally coupled theory \eqref{Action} under consideration. \\

\section{\bf{Action and Noether symmetric approach}}

In the Friedmann-Robertson-Walker minisuperspace \eqref{RW} under consideration the Ricci scalar reads as ${R}=6\big{(}\frac{\ddot{a}}{a}+\frac{\dot{a}^{2}}{a}+\frac{k}{a^{2}}\big{)}$
and therefore the action \eqref{Action} takes the following form

\be\label{A} A = \int \bigg{(}m_p^2(-3fa \dot a^2-3a^2 \dot a \dot{\phi} f'+3kaf)+\frac{1}{2}\omega \dot{\phi}^2 a^3-a^3 V-m_p^2a^3 \Lambda\bigg{)}dt \ee
in the unit $\hbar = c = 1$) while, $m_p$ ($m_p^2=\frac{1}{8\pi G}$) is the Planck mass. The above action is canonical, provided
the Hessian determinant: $W=\sum \frac{\partial^2 L}{\partial \dot a \partial \dot \phi} = -12a^4 \big{(}3{f'}^2+2\omega f\big{)}\neq 0$. The point Lagrangian is expressed (in the unit $8\pi G = 1$) as,

\be \label{Lag} L = -3fa \dot a^2-3a^2 \dot a \dot{\phi} f'+3kaf+\frac{1}{2}\omega \dot{\phi}^2 a^3-a^3 V-a^3 \Lambda. \ee
The field equations are,

\be\label{f1} \bigg{(}2\frac{\ddot a}{a}+\frac{\dot a^2}{a^2}+\frac{k}{a^2}\bigg{)}+\bigg{(}\ddot\phi+2\frac{\dot a}{a}\dot\phi\bigg{)}\frac{f'}{f}+\frac{1}{f}\bigg{(}\frac{1}{2}\dot\phi^2\omega-V\bigg{)}+\frac{f''}{f}\dot\phi^2-\frac{\Lambda}{f}=0\ee
\be\label{f2} \bigg{(}\ddot\phi+3\frac{\dot a}{a}\dot\phi\bigg{)}\frac{\omega}{3f'}-\bigg{(}\frac{\ddot a}{a}+\frac{\dot a^2}{a^2}+\frac{k}{a^2}\bigg{)}+\bigg{(}\frac{1}{2}\dot\phi^2\omega'+V'\bigg{)}\frac{1}{3f'}=0\ee
\be\label{f3} \bigg{(}\frac{\dot a^2}{a^2}+\frac{k}{a^2}\bigg{)}+\frac{\dot a}{a}\dot\phi \frac{f'}{f}-\frac{1}{3f}\bigg{(}\frac{1}{2}\dot\phi^2\omega+V\bigg{)}-\frac{\Lambda}{3f}=0\ee
where, dot denotes derivative with respect to time while prime represents derivative with respect to $\phi$. The expressions for the effective energy density ($\rho_{e}$) and the effective pressure ($p_e$) are,

\be\label{rhop} \rho_{e} = {1\over f}\left[{\omega\over 2}\dot\phi^2 + V - 3{\dot a\over a}\dot\phi f' + \Lambda\right];\\p_e = {1\over f}\left[{\omega\over 2}\dot\phi^2 - V +\left(\ddot \phi + 2{\dot a\over a}\dot\phi\right) f' + f''\dot\phi^2 - \Lambda\right].\ee
Consequently, one can also compute the sum as,

\be\label{rhop1} \rho_e+p_e = {1\over f}\left[\omega\dot\phi^2 +\left(\ddot \phi - {\dot a\over a}\dot\phi\right)f' + f''\dot\phi^2\right].\ee
As mentioned, to explore the form of the unknown parameters involved in the point Lagrangian, let us now demand Noether symmetry by imposing the condition from (\ref{Lx}) to find the following Noether equation

\be\label{xl}\begin{split}&\alpha\Big(-6a\dot a\dot\phi f'-3\dot a^2 f+3kf+\frac{3\omega}{2}\dot\phi^2 a^2-3a^2V-3\Lambda a^2\Big) +\beta\Big(-3a^2\dot a\dot\phi f''-3a\dot a^2 f'+3kaf'\\&+\frac{\omega'}{2}\dot\phi^2a^3-a^3V'\Big)
+\Big(\frac{\partial \alpha}{\partial a}\dot a + \frac{\partial\alpha}{\partial\phi}\dot\phi\Big)(-3a^2\dot\phi f'-6a\dot a f)+\Big(\frac{\partial \beta}{\partial a}\dot a+\frac{\partial\beta}{\partial\phi}\dot\phi\Big)(-3a^2\dot a f'+\omega\dot \phi a^3).\end{split}\ee
Naturally, equation (\ref{xl}) is satisfied provided the co-efficient of $\dot a^2,\dot\phi^2,\dot a\dot\phi$ and the term free from time derivative vanish separately, i.e.

\be\label{ad2}  \alpha+2a\frac{\partial\alpha}{\partial a}+a^2\frac{\partial\beta}{\partial a}\frac{f'}{f}+a\beta\frac{f'}{f} =0, \ee
\be\label{pd2}  3\alpha-6\frac{f'}{\omega}\frac{\partial\alpha}{\partial\phi}+2a\frac{\partial\beta}{\partial\phi}+a\beta\frac{\omega'}{\omega} =0, \ee
\be\label{adpd}  \big{(}2\alpha+a\frac{\partial\alpha}{\partial a}+a\frac{\partial\beta}{\partial\phi}\big{)}+a\frac{f''}{f'}\beta+2\frac{f}{f'}\frac{\partial\alpha}{\partial\phi}-\frac{\omega}{3f'} a^2\frac{\partial\beta}{\partial a} = 0, \ee
\be\label{FE}  3kf\big{(}\alpha+a\beta\frac{f'}{f}\big{)}= a^2\big{(}3V\alpha+\beta V'a+3\Lambda\alpha\big{)}.\ee
We now look for the conditions on the integrability of this set of above equations \eqref{ad2} - \eqref{FE}. Since, here the number of equations are four while the number of unknown parameters are five $(\alpha, \beta, \omega, f, V)$, so the set of above equations can not be solved exactly unless extra condition is imposed. Rather, we obtain restrictions on the forms of $\alpha$, $\beta$, $f$, $\omega$ and $V$. This will leave large freedom of choice, so that all the interesting cases may be accommodated. However, as mentioned, we shall in the present manuscript restrict ourselves to study only three cases of particular importance. The set of particular differential equations are solved under the assumption that $\alpha$ and $\beta$ are separable (and non null), i.e.

\be\label{alphabeta} \alpha(a,\phi)=A_1(a)B_1(\phi);\\  \beta(a,\phi)=A_2(a)B_2(\phi).\ee
With these assumptions, the integrability conditions are (See Appendix)

\be\label{A12} A_1=-\frac{cl}{a};\\B_2=c\frac{f}{f'}B_1;\\A_2=\frac{l}{a^2};\\V=V_o
f^3-\Lambda;\\B'_1=\frac{2}{3}\frac{\omega}{f'}B_1;\\ 3f'^2+2\omega f=\frac{n}{4}\omega f^3\ee
where, $c,l,n,V_0$ are all arbitrary constants. Note that since $f(\phi) \ne 0$ and $\omega(\phi) \ne 0$, so the nondegeneracy condition remains satisfied provided $n \ne 0$. Clearly, we need to solve the above six equation \eqref{A12}, for seven unknowns, viz. $(A_1,A_2,B_1,B_2,f,V,\omega)$. It is important to mention that while general conserved current always exists for $V \varpropto f^2$ \cite{47, 48, 49, 58, 59}, Noether symmetry exists for $V \varpropto f^3$, in the absence of Lambda. This clearly depicts that Noether symmetry procedure is unable to explore all the available symmetries of a theory. For $\omega=1$, a general nonminimally coupled case we get in view. The last relation of equation (\ref{A12}) then gives an elliptic integral, which can be solved for $f$ in closed form only under the assumption $n=0$. But this makes the Hessian determinant $W=0$, and so the Lagrangian turns out to be degenerate. Also, for $n = 0 $, the general solution of (\ref{A12}) is, $\label{n=0} f=-\frac{(\phi-\phi_0)^2}{6}$, which makes the Newtonian gravitational constant $G$ negative \cite{42}. Thus we omit the case $\omega = 1$.

\section{\bf{Solutions under different choice of the coupling parameter $f(\phi)$}}

Let us make things clear yet again, for a consistency check. To get a picture of evolution of the early universe in view of the action \eqref{Action}, we need to solve the set of Einstein's field equations \eqref{f1}, \eqref{f2} and \eqref{f3} exactly. Out of which only two are independent and they involve $5$ unknowns ($a(t), \phi(t), f(\phi), \omega(\phi), V(\phi)$) altogether. Clearly, one requires $3$ physically reasonable assumptions for the purpose, and the standard followup is to choose some typical forms of $f(\phi), \omega(\phi)$ and $V(\phi)$. Instead we impose Noether symmetry i.e. $\pounds_X L = 0$, as our first assumption, since nothing is more physical in the world than symmetry. As a result we find four equations \eqref{ad2}, \eqref{pd2}, \eqref{adpd} and \eqref{FE}, with five unknown parameters viz. $\alpha, \beta, f, \omega, V$. Thus at this stage we require just one more assumption to exactly solve the above set of Noether equations. To handle the above set of partial differential equations \eqref{ad2} through to \eqref{FE} we consider separation of variables and ended up with yet another set of equations \eqref{A12}. One can clearly notice that already $A_1$ and $A_2$ are found exactly as functions of the scale factor $a(t)$, while we are left with four relations in \eqref{A12} with five parameters $B_1, B_2, f, \omega, V$. Hence still one needs one more assumption to explicitly find the forms of these five parameters. This proves everything is consistent so far. One can generate indefinitely large number of symmetries and hence exact cosmological solutions, by making different choices of one of the parameters. In this section however, we shall study only three different cases making reasonable assumptions on three different forms of $f(\phi)$, since as already known, different forms of $f(\phi)$ leads to different physical theory. The three cases represent `Brans-Dicke theory of gravity', `Induced theory of gravity' and `General non-minimal theory of gravity'. As a result we find $\alpha(a, \phi)$, $\beta(a,\phi)$, $\omega(\phi)$ and $V(\phi)$, and hence the conserved current. We shall then express the Lagrangian in terms of the new variables $u$ and $v$, $u$ being cyclic, and use the conserved current and the energy equation expressed in terms of the new variables as

\be\label{QE} Q = {\partial L\over \partial \dot u};\\E_L = \frac{\partial L}{\partial\dot u}\dot u+\frac{\partial L}{\partial\dot v}\dot v-L = 0,\ee
to solve the Einstein's field equations exactly in all the three different cases.

\paragraph{\textbf{Case 1. Brans-Dicke theory.}} First we consider the well-known Brans-Dicke theory by choosing $f(\phi) = \phi$. In view of Equation \eqref{A12}, we therefore obtain the following solutions

\be \label{V1} V=V_0\phi^3-\Lambda, \\ \omega=\frac{12}{n\phi^3-8\phi},\\ B_1=B_{0}\frac{\sqrt{n\phi^2-8}}{\phi},\\ B_2=cB_{0}\sqrt{n\phi^2-8},\ee
where, $B_{0}$ is yet another constant. In view of equations \eqref{A12} and \eqref{V1} $\alpha$ and $\beta$ are obtained as,

\be \alpha=-C\frac{\sqrt{n\phi^2-8}}{a\phi},\\\beta=C\frac{\sqrt{n\phi^2-8}}{a^2},\ee
where the constant $C = c_0B_0l$. So the conserved current \eqref{Q} is found as,

\be \label{Q1} Q=3Ca\sqrt{n\phi^2-8}\bigg{(}\frac{\dot a}{a}+\frac{n\phi^2-4}{n\phi^2-8}\frac{\dot\phi}{\phi}\bigg{)}.\ee
Using the forms of $f(\phi) = \phi$ and the forms of $\omega(\phi), ~V(\phi)$ presented in equation \eqref{V1}, the point Lagrangian \eqref{Lag} may now be expressed as,

\be\label{Lag1} L=-3a\dot a^2\phi-3a^2\dot a\dot\phi+3ka\phi+\frac{6a^3 \dot\phi^2}{n\phi^3-8\phi}-V_0 a^3\phi^3.\ee
At this stage let us perform the change of variables to obtain the corresponding cyclic coordinate associated with the conserved current \eqref{Q1}. Equation (\ref{cyclic}) is solved exactly under the following choice,

\be \label{u1} u=\frac{a^2\phi}{8}\sqrt{n\phi^2-8};\\v=a\phi,\ee
which may be inverted to yield

\be a^2=\frac{nv^4-64u^2}{8v^2};\\ \phi^2=\frac{8v^4}{nv^4-64u^2}.\ee
Being always $a > 0$, the Jacobian of transformation does not give any trouble, and the same holds for all the cases studied below. Under the transformation \eqref{u1}, the Lagrangian \eqref{Lag1} takes the form

\be L=6\frac{\dot u^2}{v}-\frac{3}{8}nv\dot v^2+3kv-V_0v^3, \ee
u being cyclic, the conserved current \eqref{QE} reads as,

\be\label{q1} Q = \frac{12\dot u}{v},\ee
and the energy equation \eqref{QE} leads to the following first order differential equation for $v$,

\be \left(\frac{Q^2}{24}-3k\right)+V_0v^2=\frac{3n}{8}\dot v^2,\ee
which may be integrated to obtain the following solution for $v$ as,

\be\label{v1} v= \frac{e^{pt}+4FV_0e^{-pt}}{4V_0},\ee
where, $F=3k-\frac{Q^2}{24}$. We may also obtain in view of equation \eqref{q1}, the form of the cyclic coordinate $u$ as,

\be\label{u11} u=\bigg{(}\frac{\sqrt{n}Q}{32\sqrt{6}V_0^{\frac{3}{2}}}\bigg{)}\bigg{(}e^{pt}-4FV_0e^{-pt}\bigg{)}+u_0\ee
where, $p=\sqrt{\frac{8V_0}{3n}}$. Setting the integration constant $u_0=0$, for the origin of time, the exact solution for $a(t)$ and $\phi(t)$ are found as

\be a(t)=\bigg{(}\frac{\frac{n}{8}(a_1 e^{pt}+a_2 e^{-pt})^4-8(a_3e^{pt}-a_4e^{-pt})^2}{(a_1e^{pt}+a_2e^{-pt})^2}\bigg{)}^{\frac{1}{2}},\ee \be\phi(t)=\frac{(a_1e^{pt}+a_2e^{-pt})^2}{\Big{(}\frac{n}{8}(a_1 e^{pt}+a_2 e^{-pt})^4-8(a_3e^{pt}-a_4e^{-pt})^2\Big{)}^{\frac{1}{2}}},\ee
Where, $a_1=\frac{1}{4V_0}$, $a_2=3k-\frac{Q^2}{24}$, $a_3=\frac{\sqrt{n}Q}{32\sqrt{6}V_0^{\frac{3}{2}}}$ and
$a_4=\frac{\sqrt{n}}{8\sqrt{6V_0}} Q\Big{(}3k-\frac{Q^2}{24}\Big{)}$ are constants. As $t \rightarrow \infty$, the scale factor $a \rightarrow {\sqrt {n\over 8}} a_1 e^{pt}$, while as $t \rightarrow -\infty$, the scale factor $a \rightarrow {\sqrt {n\over 8}} a_2 e^{pt}$, and finally as $t \rightarrow 0$, the scale factor $a \rightarrow \sqrt {{n\over 8} (a_1+a_2)^2 - 8\big({a_3-a_4\over a_1+a_2}\big)^2}$. Therefore asymptotically ($t\rightarrow \pm \infty$) the universe is de-Sitter, while as $t \rightarrow 0$, it has a finite radius. Therefore the solution represents Lorentzian wormhole. One can also make a further check in a straightforward manner. Asymptotically ($t \rightarrow \pm \infty$), the scalar field turns out to be a constant $\phi \rightarrow \sqrt{8\over n}$. As a result in the present case in view of equations \eqref{rhop} and \eqref{rhop1}, asymptotically one finds $\rho_e \rightarrow V_0\phi^2 > 0$, $p_e \rightarrow -V_0\phi^2$ and $\rho_e + p_e \rightarrow 0$. Thus the WEC is satisfied, and simultaneously asymptotic de-Sitter expansion is confirmed. In the process, the present solution also is in tune with the fact mentioned in the introduction that, for Einstein gravity there are non-static Lorentzian wormholes which do not require WEC violating matter to sustain them \cite{23, 24}.

\paragraph{\textbf{Case 2. Induced theory of gravity}} Let us now consider induced theory of gravity by the choice $f(\phi)=\epsilon\phi^2$, where $\epsilon$ is the coupling constant. Under this choice, we obtain the following solutions in view of Equation (\ref{A12}),

\be \label{V2} V=V_0  \epsilon^3\phi^6-\Lambda, \\ \omega=\frac{48\epsilon}{n\epsilon^2\phi^4-8},\\ B_1=D_0\sqrt{\frac{n\epsilon^2\phi^4-8}{n\epsilon^2\phi^4}}, \\B_2=cD_0\sqrt{\frac{n\epsilon^2\phi^4-8}{4n\epsilon^2\phi^2}},\ee
where $c,D_0$ are constant. As a result we also find

\be\alpha=-\frac{N\sqrt{n\epsilon^2\phi^4-8}}{a\phi^2},\\ \beta=\frac{N\sqrt{n\epsilon^2\phi^4-8}}{2a^2\phi},\ee
where the constant $N=\frac{cD_0 l}{\epsilon\sqrt{n}}$. The conserved current in the present case reads as,

\be Q=3N\epsilon a\sqrt{n\epsilon^2\phi^4-8}\bigg{(}\frac{\dot a}{a}+2\frac{n\epsilon^2\phi^4-4}{n\epsilon^2\phi^4-8}\frac{\dot\phi}{\phi}\bigg{)}.\ee
The Lagrangian \eqref{Lag} takes the form,

\be\label{Lag2} L=-3\epsilon a\dot a^2\phi^2-6\epsilon a^2\dot a\phi\dot\phi+3\epsilon ka\phi^2+\frac{24\epsilon a^3 \dot\phi^2}{n\epsilon ^2\phi^4-8}-V_0\epsilon^3 a^3\phi^6.\ee
As before, let us now perform the change of variables to obtain the corresponding cyclic coordinate $u$. The equation \eqref{cyclic} is satisfied under the choice

\be u=\frac{a^2\phi^2}{8}\sqrt{n\epsilon^2\phi^4-8};\\v=a\phi^2,\ee
which may be inverted to obtain

\be a^2=\frac{n\epsilon^2v^4-64u^2}{8v^2},\\ \phi^2=\frac{\sqrt{8v^4}}{\sqrt{n\epsilon^2v^4-64u^2}},\ee
while the Lagrangian \eqref{Lag2} in view of the new variables now takes the following form,

\be L=\frac{6\epsilon\dot u^2}{v}-\frac{3}{8}n\epsilon^3v\dot v^2+3\epsilon kv-V_0 \epsilon^3v^3. \ee
Now, u being cyclic, the conserved current \eqref{QE} reads as,

\be \label{q2} Q = \frac{12\epsilon\dot u}{v}.\ee
In view of the energy equation \eqref{QE} we find the following first order differential equation in $v$,

\be \Big(\frac{Q^2}{24\epsilon^4}-\frac{3k}{\epsilon^2}\Big)+V_0v^2=\frac{3n}{8}\dot v^2.\ee
The above first order differential equation in $v$ may be integrated to find the following form of $v$,

\be\label{v1} v= \frac{e^{pt}+4F_0V_0e^{-pt}}{4V_0},\ee
where, $F_0=\frac{3k}{\epsilon^2}-\frac{Q^2}{24\epsilon^4}$. The cyclic variable $u$ may be found as well in view of \eqref{q2} as,

\be\label{u} u=\bigg{(}\frac{Q\sqrt{n}}{32V_0\epsilon\sqrt{6V_0}}\bigg{)}\bigg{(}e^{pt}-4F_0V_0e^{-pt}\bigg{)}+u_0,\ee
where, $p=\sqrt{\frac{8V_0}{3n}}$. Setting the integration constant $u_0 = 0$ as before, we finally obtain exact solutions of $a(t)$ and $\phi(t)$ as,

\be a(t)=\bigg{(}\frac{\frac{n\epsilon^2}{8}(b_1 e^{pt}+b_2 e^{-pt})^4-8(b_3e^{pt}-b_4e^{-pt})^2}{(b_1e^{pt}+b_2e^{-pt})^2}\bigg{)}^{\frac{1}{2}},\ee
\be \phi(t)=\frac{(b_1e^{pt}+b_2e^{-pt})}{\bigg{(}\frac{n\epsilon^2}{8}(b_1 e^{pt}+b_2 e^{-pt})^4-8(b_3e^{pt}-b_4e^{-pt})^2\bigg{)}^{\frac{1}{4}}},\ee
where, $b_1=\frac{1}{4V_0}$,~$b_2=\frac{3k}{\epsilon^2}-\frac{Q^2}{24\epsilon^4}$,~$b_3=\frac{Q\sqrt{n}}{32V_0\epsilon\sqrt{6V_0}}$,
$b_4=\big{(}\frac{3k}{\epsilon^2}-\frac{Q^2}{24\epsilon^4}\big{)}\big{(}\frac{Q\sqrt{n}}{8\epsilon\sqrt{6V_0}}\big{)}$ are constants as specified. Clearly, the form of the scale factor remains unaltered from the previous case, and as such represents Lorentzian wormhole solution yet again. Weak energy condition $\rho_e > 0$ and $\rho_e + p_e \ge 0$ is satisfied here too.

\paragraph{\textbf{Case 3. Non-minimally coupled theory of gravity.}} Finally, let us consider $f(\phi) = (1-\varepsilon\phi^2)$, $\varepsilon$ being a coupling constant. Under this choice, Equation \eqref{A12} yields the following set of solutions,

\be V=V_0(1-\varepsilon\phi^2)^3-\Lambda, \\ \omega=\frac{48\varepsilon^2\phi^2}{(1-\varepsilon\phi^2)[n(1-\varepsilon \phi^2)^2-8]},\\B_1=B_0\frac{\sqrt{n(1-\varepsilon \phi^2)^2-8}}{\sqrt{n}(1-\varepsilon\phi^2)}, \\B_2=-cB_0\frac{\sqrt{n(1-\varepsilon \phi^2)^2-8}}{2\sqrt{n}\varepsilon\phi},\ee
where $c,B_0$ are constants. As a result we find

\be\alpha=-\frac{N_0\sqrt{n(1-\varepsilon \phi^2)^2-8}}{a(1-\varepsilon\phi^2)},\\ \beta=-\frac{N_0\sqrt{n(1-\varepsilon \phi^2)^2-8}}{2\varepsilon a^2\phi},\ee
where $N_0=\frac{cB_0 l}{\sqrt{n}}$ is a constant. The conserved current turns out to be,

\be Q=3N_0 a\sqrt{n(1-\varepsilon \phi^2)^2-8}\bigg{(}\frac{\dot a}{a}+2\varepsilon\frac{\phi\dot\phi}{1-\varepsilon\phi^2}\frac{n(1-\varepsilon \phi^2)^2-4}{n(1-\varepsilon \phi^2)^2-8}\bigg{)},\ee
while the Lagrangian \eqref{Lag} takes the following form,

\be\label{Lag3} L=-3a\dot a^2(1-\varepsilon\phi^2)+6\varepsilon a^2\dot a\phi \dot\phi+3ka(1-\varepsilon\phi^2)+\frac{24 \varepsilon^2 a^3 \phi^2\dot\phi^2}{(1-\varepsilon\phi^2)\big{(}n(1-\varepsilon\phi^2)-8\big{)}}-V_0 a^3(1-\varepsilon\phi^2)^3.\ee
As before, we now perform the change of variables to obtain the corresponding cyclic coordinate $u$. Equation \eqref{cyclic} may be solved to find,

\be u=\frac{a^2(1-\varepsilon\phi^2)}{8}\sqrt{n(1-\varepsilon\phi^2)^2-8};\\v=a(1-\varepsilon\phi^2),\ee
which may be inverted to obtain

\be a^2=\frac{nv^4-64u^2}{8v^2}, \\ \phi^2=\frac{1}{\varepsilon}\bigg{(}1-\frac{8v^4}{nv^4-64u^2}\bigg{)}.\ee
The Lagrangian \eqref{Lag3} in terms of the new variables ($u$ and $v$) takes the following simplified form

\be L=\frac{6\dot u^2}{v}-\frac{3}{8}n v\dot v^2+3 kv-V_0v^3.\ee
Since u is cyclic, the conserved current may be found in view of \eqref{QE} as,

\be \label{q3} Q = \frac{12\dot u}{v},\ee
while the energy equation \eqref{QE} reads as,

\be \Big(\frac{Q^2}{24}-3k\Big)+V_0v^2=\frac{3n}{8}\dot v^2,\ee
which may be integrated to yield

\be\label{v1} v= \frac{e^{pt}+4FV_0e^{-pt}}{4V_0},\ee
where, $F=3k-\frac{Q^2}{24}$. We can also find the cyclic coordinate $u$ in view of \eqref{q3} as,

\be\label{u} u=\bigg{(}\frac{\sqrt{n}Q}{32\sqrt{6}V_0^{\frac{3}{2}}}\bigg{)}\bigg{(}e^{pt}-4FV_0e^{-pt}\bigg{)}+u_0,\ee
where $p=\sqrt{\frac{8V_0}{3n}}$. As before we set the integration constant $u_0$ to zero for the origin of time, and solve the scale factor $a(t)$ and the scalar field $\phi(t)$ exactly as

\be a(t)=\bigg{(}\frac{\frac{n}{8}(a_1 e^{pt}+a_2 e^{-pt})^4-8(a_3e^{pt}-a_4e^{-pt})^2}{(a_1e^{pt}+a_2e^{-pt})^2}\bigg{)}^{\frac{1}{2}},\ee \be\phi(t)=\frac{1}{\sqrt{\varepsilon}}\bigg{(}1-\frac{(a_1e^{pt}+a_2e^{-pt})^4}{\frac{n}{8}(a_1 e^{pt}+a_2 e^{-pt})^4-8(a_3e^{pt}-a_4e^{-pt})^2}\bigg{)}^{\frac{1}{2}},\ee
where, $a_1=\frac{1}{4V_0}$,~$a_2=3k-\frac{Q^2}{24}$,~$a_3=\frac{\sqrt{n}Q}{32\sqrt{6}V_0^{\frac{3}{2}}}$, and
$a_4=\frac{\sqrt{n}}{8\sqrt{6V_0}} Q\bigg{(}3k-\frac{Q^2}{24}\bigg{)}$ are constants. Here again we observe that the form of the scale factor remains unaltered from the earlier ones and therefore represents Lorentzian wormhole solution. The weak energy condition is not violated here again.

\section{\bf{Discussion and conclusions}}

Excitement raised after Ruggiero et al \cite{45, 46} for the first time applied Noether symmetry in the scalar-tensor theory of gravity, to find a form of potential which naturally led to `Inflation'. Thereafter, many people worked in the field and proved it to be a very powerful tool to explore the parameters and the potential involved in a theory. It also make things much easier to solve the Einstein's field equations, particularly in view of the cyclic coordinate. The technique has been applied here again for a general non-minimally coupled scalar-tensor theory of gravity, in the presence of cosmological constant. Three cases of particular interest have been studied, viz. the `Brans-Dicke theory', the `Induced theory of gravity' and the `General non-minimal scalar-tensor theory of gravity'. While only an imaginary scalar field admits classical Euclidean wormhole solution \cite{8}, here, its Lorentzian counterpart admits wormhole solutions for real scalar field. Noticeably, all the cases having different coupling parameters and potentials yield the same form of the scale factor, which represents cosmological Lorentzian wormholes. Such wormholes admit weak energy condition. While Euclidian wormholes do not exist in general for arbitrary potential \cite{8}, evolving cosmological Lorentzian wormholes on the contrary, appear to be a generic feature of non-minimally coupled Scalar-tensor theory of gravity. Such solutions depicts that, the universe itself evolved as a Lorentzian wormhole, which initiates inflation thereafter.In the process, it removes cosmological singularity arising from GTR even at the classical level. It is now required to check if the inflationary behaviour is at par with the currently released data \cite{60}, which we pose in future.

\appendix
\section{\bf {Appendix}}
In the appendix we explicitly solve the set of Noether equations \eqref{ad2} - \eqref{FE}, under separation of variables \eqref{alphabeta}, to demonstrate that the solutions lead to the set of equation \eqref{A12}. Primarily, equation \eqref{ad2} takes the form

\be\label{C1} \frac{A_1+2aA'_1}{a(A_2+aA'_2)}=-\frac{B_2}{B_1}\frac{f'}{f}=-C_1,\ee
under the condition $a(A_2+aA'_2)\neq0$. Next, equation \eqref{pd2} takes the form

\be\label{C2} \frac{3B_1-6\frac{f'}{\omega}B'_1}{2B'_2+\frac{\omega'}{\omega}B_2}=-a\frac{A_2}{A_1}=-C_2,\ee
and of-course we need to fix $({2B'_2+\frac{\omega'}{\omega}B_2})\neq 0$.
Now, using the relations ({\ref{C1}}) and ({\ref{C2}}), we get the following relation from equation (\ref{FE})

\be 3kf(1+C_1C_2)=a^2\left(3V+V'C_1C_2\frac{f}{f'}+3\Lambda\right),\ee
which, for $k\neq0$ implies that:

\be\label{c1c2} C_1C_2=-1,\\ V=V_0 f^3-\Lambda.\ee
In order to obtain $A_1$ and $A_2$, let us set $C_1 = c = − \frac{1}{C_2}$, and use it in equations (\ref{C1}) and (\ref{C2}). As a result, we obtain,

\be\label{A2} A_2=\frac{l}{a^2},\\ A_1=-\frac{cl}{a},\\ B_2=c\frac{f}{f'}B_1.\ee
Finally, using the equations (\ref{adpd}) and (\ref{C2}), we obtain the last two relations appearing in (\ref{A12}), viz.

\be\label{B'1} B'_1=\frac{2\omega}{3f'}B_1,\\3f'^2+2\omega f=\frac{n}{4}\omega f^3.\ee
It is interesting to note that the equations (\ref{A2}) and (\ref{B'1}) naturally lead to the following general relation between $\alpha(a,\phi)$ and $\beta(a,\phi)$, viz.

\be\label{alpa=beta} \alpha=-a\beta\frac{f'}{f}.\ee

\end{document}